\documentclass[prd,twocolumn,floatfix,preprintnumbers,letterpaper]{revtex4}
\usepackage{graphicx}
\usepackage{amsmath,amssymb}
\input{epsf}
\usepackage{epsf}

\begin{document}

\title{Dark Energy Models in the $w - w^\prime$ Plane}
\author{Robert J. Scherrer}
\affiliation{Department of Physics and Astronomy, Vanderbilt University,
Nashville, TN  ~~37235}

\begin{abstract}
We examine the behavior of dark energy models in the plane defined
by $w$ (the equation of state parameter for the dark energy) and $w^\prime$
(the derivative
of $w$ with respect to the logarithm of the scale factor).
For non-phantom barotropic fluids with positive squared sound speed, we find
that $w^\prime < 3w(w+1)$, the opposite of the bound on
quintessence models previously derived by Caldwell and Linder.
Thus, these barotropic models and quintessence models for the dark
energy occupy disjoint regions in the $w-w^\prime$ plane.
We also derive two new bounds for quintessence models in the
$w - w^\prime$ plane: the first is a general
bound for any scalar field with a monotonic potential,
while the second improves on the Caldwell-Linder bound
for tracker quintessence models.  Observationally
distinguishing barotropic
models from quintessence models requires
$\sigma(w^\prime) \lesssim 1+w$.

\end{abstract}

\maketitle

\section{Introduction}

Observations indicate that roughly
70\% of the total energy density in the universe is in the form
of dark
energy, which has negative
pressure and drives the late-time acceleration 
 of the universe (see Ref.
\cite{Sahni} for a recent review, and references therein).

The simplest model for this dark energy is a cosmological constant,
but many other alternatives have been put forward.  One possibility,
dubbed quintessence, is a model in which the dark energy
arises from a scalar
field \cite{ratra,turner,caldwelletal,liddle,zlatev}.
In another class of models,
the dark energy is simply
taken to be  barotropic
fluid, in which the pressure $p$ and energy density $\rho$ are related by
\begin{equation}
p = f(\rho).
\end{equation}
The prototype for this sort of model is
the Chaplygin gas \cite{Kamenshchik,Bilic},
which can be considered a special case of the generalized Chaplygin gas
\cite{Bento}.
Although originally proposed as unified models for dark matter and dark
energy, the Chaplygin gas and generalized Chaplygin gas models
have also been examined as models for dark energy alone
\cite{Dev,Gorini,Bean,Mul,Sen}.
Interest in barotropic fluids as dark energy has not been
confined to the generalized Chaplygin gas.  Many 
other models have been proposed, including, e.g.,  the Van der Waals
model \cite{VDW} and the wet dark fluid model \cite{water}.

Obviously, it is important to determine whether these models
can be distinguished by their observational consequences.  In
this vein, Caldwell and Linder \cite{CL} examined the behavior
of quintessence models in the plane defined by the quantities
$w$ and $w^\prime$ for the dark energy.  Here $w$ is the ratio of
pressure to density for the dark energy:
\begin{equation}
w = p/\rho,
\end{equation}
and $w^\prime$ is the derivative of $w$ with respect to the logarithm
of the scale factor $a$:
\begin{equation}
w^\prime = \frac{dw}{d\ln(a)}.
\end{equation}
Caldwell and Linder showed that quintessence models in which the
scalar field potential asymptotically approaches zero
can be divided naturally into two categories, which they dubbed
``freezing" and ``thawing" models, with quite different
behavior in the $w- w^\prime$ plane.

Here we extend and generalize these results.  In the next section,
we determine the behavior of general non-phantom barotropic fluids with positive
squared sound speed in the $w - w^\prime$ plane, and we show
that these models occupy a region of phase space which is disjoint
from that occupied by quintessence models.  In Section 3, we
re-examine the lower bound on $w^\prime$ from Ref. \cite{CL};
we derive a new bound for
a more general class of quintessence models,
and sharpen the bound from Ref. \cite {CL} for the case of tracker potentials.
These results are discussed in Section 4.
Our results support the argument
that the behavior of dark energy models
in the $w - w^\prime$ plane provides a useful discriminator for
such models.

\section{Barotropic Fluids}

As noted in the previous section, a barotropic fluid is one for which
the pressure is purely a function of the density, $p = f(\rho)$.
For such models, we have
\begin{equation}
w^\prime = \frac{dw}{d\rho} \frac{d\rho}{d\ln(a)}. 
\end{equation}
The factors in this equation are given by
\begin{equation}
\frac{dw}{d\rho} = \frac{1}{\rho}\left(\frac{dp}{d\rho} - w\right),
\end{equation}
and
\begin{equation} 
\frac{d\rho}{d\ln(a)} = -3(1+w)\rho.
\end{equation}

Then the expression for $w^\prime$ becomes
\begin{equation}
\label{baro}
w^\prime = -3(1+w)\left(\frac{dp}{d\rho} - w\right).
\end{equation}
In order to derive useful constraints from equation
(\ref{baro}), we need to assume something about
the behavior of $dp/d\rho$ and $w$.  We will assume
first that we are not dealing with phantom models,
so that $1+w >0$.  Second, we note that $dp/d\rho = c_s^2$,
and we will assume further that $c_s^2 > 0$, to
avoid instabilities in perturbation growth.  These
constraints are satisfied by most barotropic fluids
of interest, although it is also possible
to construct models which violate both assumptions (see,
e.g., Ref. \cite{Sen}).  With these assumptions,
equation (\ref{baro}) gives the limit
\begin{equation}
\label{barobound}
w^\prime < 3w(1+w).
\end{equation}
This limit is shown in Fig. 1, where have used the
same scale as Ref. \cite{CL} for easy comparison.
\begin{figure}[htb]
\centerline{\epsfxsize=3truein\epsffile{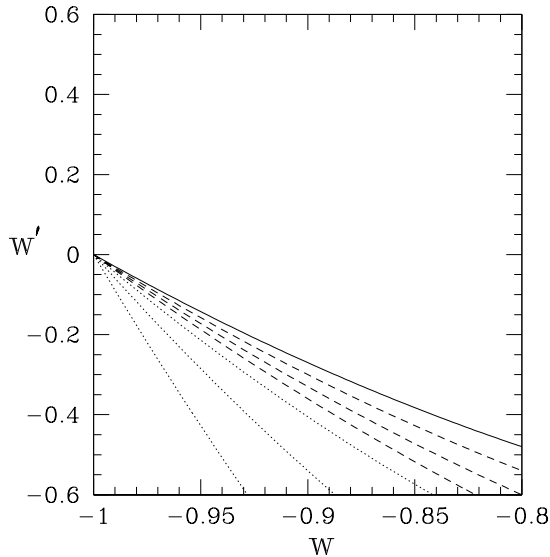}}
\caption{Solid curve gives the upper bound on $w^\prime$ as a function
of $w$ for non-phantom barotropic models with positive squared sound
speed.  Dotted curves give $w^\prime$ as a function of $w$ for
Chaplygin gas dark energy with (top to bottom) $\alpha = 0.5$, $\alpha=1$,
$\alpha = 2$,
while $\alpha=0$ lies on the solid curve.  Dashed curves give $w^\prime$
as a function of $w$ for the wet dark fluid model with (top to bottom)
$w_* = 0.1$, $w_* = 0.2$, $w_* = 0.3$.}
\end{figure}
Note that this is the exact opposite of the limit
derived in Ref. \cite{CL} for ``freezing" quintessence models:
\begin{equation}
w^\prime > 3w(1+w).
\end{equation}
Thus, barotropic fluids and freezing quintessence models occupy non-overlapping
regions in the $w - w^\prime$ plane.  The bound in
equation (\ref{barobound}) is saturated when $c_s^2 = dp/d\rho = 0$,
which corresponds to the case $p = constant$.  In this
case, we have $w^\prime = 3w(1+w)$.  Models of this kind
are particularly interesting, since they correspond
to a barotropic fluid that behaves exactly
like a mixture of dark matter (dust) and a cosmological constant.
Such models can be constructed in the context of either
the generalized Chaplygin gas \cite{avelino} or $k$-essence \cite{kess}.

Now consider several illustrative special cases.  For the
generalized Chaplygin gas, the
equation of state is given by
\begin{equation}
\label{Chap}
p = -\frac{A}{\rho^{\alpha}},
\end{equation}
where $A$ and $\alpha$ are constants.  (The Chaplygin gas
corresponds to the special case $\alpha = 1$).
Then it is straightforward
to use equation (\ref{baro}) to derive
\begin{equation}
w^\prime = 3(1+\alpha)w(1+w).
\end{equation}
This behavior is displayed in Fig. 1 for several values of $\alpha$.
The $\alpha=0$ case saturates the upper bound on $w^\prime$, as it corresponds
to the previously-noted $p = constant$ model, while
the original Chaplygin gas model ($\alpha =1$) gives
$w^\prime = 6w(1+w)$.

Another barotropic model is the wet dark fluid model of
Ref. \cite{water}, with equation of state
\begin{equation}
p = w_*(\rho - \rho_*),
\end{equation}
where $w_*$ and $\rho_*$ are constants.  Again, it is straightforward
to use equation (\ref{baro}) to derive the result
\begin{equation}
w^\prime = -3(1+w)(w_* - w),
\end{equation}
which is displayed in Fig. 1 for several values of $w_*$.

While we have assumed
that the fluid characterized by $w^\prime$ and $w$
acts strictly as dark energy, all of the results derived here
can also be applied to unified dark energy models, in which
the barotropic fluid serves as both dark matter and dark energy
(indeed, this is one of the main motivations for the Chaplygin
gas model and its variants).  However, in the
case of unified dark energy, $w$ and $w^\prime$ refer
to the entire fluid, not just to the dark energy component.

\section{Quintessence}

The equation of motion for the quintessence field $\phi$ is
\begin{equation}
\ddot\phi + 3H\dot\phi + \frac{dV}{d\phi} = 0,
\end{equation}
where the Hubble parameter $H$ is
\begin{equation}
H = \frac{\dot a}{a} = \left(\frac{\rho}{3}\right)^{1/2},
\end{equation}
and $\rho$ is the total density.
(We assume an $\Omega=1$ universe and take $8 \pi G = 1$ throughout).
Since we are interested in the late-time evolution of
the dark energy, we assume that $\rho = \rho_M + \rho_\phi$, where
$\rho_M$ is the density of the matter, scaling as $\rho_M \propto a^{-3}$, and
$\rho_\phi$ is the scalar field energy density, given by
\begin{equation}
\rho_\phi = \frac{\dot \phi^2}{2} + V(\phi).
\end{equation}
The pressure of the scalar field is
\begin{equation}
p_\phi = \frac{\dot \phi^2}{2} - V(\phi).
\end{equation}

Caldwell and Linder \cite{CL} examined models in which $\phi$ evolves
toward the state $V(\phi) = 0$.  They distinguished two cases:
``thawing" models, in which $w \approx -1$ initially, but $w$ increases
as $\phi$ rolls down the potential, and ``freezing" models, in which
$w > -1$ initially, but $w$ approaches $-1$ as the field rolls
down the potential.  The bounds derived in Ref. \cite{CL} are
based on a combination of plausibility arguments and numerical simulations.
For the ``thawing" models, these limits are
\begin{equation}
1+w < w^\prime < 3(1+w),
\end{equation}
while for the ``freezing" models,
\begin{equation}
\label{freeze}
3w(1+w) < w^\prime < 0.2w(1+w).
\end{equation}

As noted in the previous section, the allowed region for barotropic
fluids in the $w - w^\prime$ plane lies adjacent to the lower
bound on $w^\prime$ for quintessence models proposed in Ref. \cite{CL}.
Therefore, it is the lower bound in equation (\ref{freeze}) which is
of greatest interest here.

We first generalize the Caldwell-Linder lower bound to a wider
range of potentials.  Consider a scalar field evolving in an
arbitrary monotonic potential $V(\phi)$, not necessarily asymptotically
zero.  We will take $V^\prime(\phi) < 0$, but of course the argument
is identical in the opposite case.  Now, following
Steinhardt et al. \cite{zlatev}, we note that
the equation of motion for $\phi$ can be rewritten as
\begin{equation}
\label{motion1}
-\frac{V^\prime}{V} = \sqrt{\frac{3(1+w)}{\Omega_\phi}}\left[
1 + \frac{X^\prime}{6}\right],
\end{equation}
where $X$
is defined as
\begin{equation}
X = \ln\left(\frac{1+w}{1-w}\right),
\end{equation}
and $X^\prime$ is the derivative of $X$ with respect to $ln(a)$,
so that $X^\prime$ and $w^\prime$ are related via
\begin{equation}
\label{xprime}
X^\prime = \frac{2 w^\prime}{(1-w)(1+w)}.
\end{equation}

Now we note that for a scalar field
rolling downhill in a monotonically decreasing potential,
the left-hand side of equation (\ref{motion1}) is always positive.
This, in turn, implies that $1 + X^\prime/6 > 0$, which translates
into the following bound on $w^\prime$:
\begin{equation}
\label{bound1}
w^\prime > -3(1-w)(1+w).
\end{equation}
While not as tight as the Caldwell-Linder bound, equation (\ref{bound1})
applies to a more general class of models; it assumes nothing about tracking
behavior or ``freezing" of the scalar field, nor is it confined
to potentials for which $V(\phi) \rightarrow 0$ asymptotically.
This bound is saturated by a scalar field rolling on the constant
potential $V = V_0$.  For this case, $w$ evolves from $w\approx +1$ to
$w \approx -1$, with $w^\prime = -3(1-w)(1+w)$ \cite{Linder}.  The bound
in equation (\ref{bound1}), along with the lower bound on $w^\prime$
from Ref. \cite{CL}, are shown in Fig. 2.
\begin{figure}[htb]
\centerline{\epsfxsize=3truein\epsffile{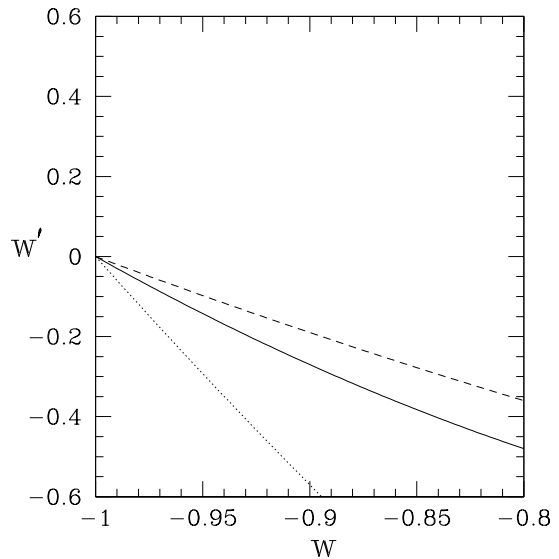}}
\caption{Solid curve gives the lower bound on $w^\prime$ as a function
of $w$ for ``freezing" quintessence models
from Ref. \cite{CL}.  Dotted curve gives our lower bound on
$w^\prime$ as a function of $w$ for quintessence
models with an arbitrary monotonic potential (equation \ref{bound1}).
Dashed curve gives our lower bound on $w^\prime$ as a function
of $w$ for tracker quintessence models (equation \ref{wprime2}).}
\end{figure}

We can tighten this bound considerably for the case of ``tracker" models,
i.e., models for which $w$ is initially constant when $\rho_\phi \ll \rho_M$,
but which evolve toward $w = -1$ when $\rho_{\phi} \approx \rho_M$.
Again, following Ref. \cite{zlatev}, we 
define $\Gamma$ to be given by
\begin{equation}
\Gamma \equiv \frac{V^{\prime\prime}V}{V^{\prime 2}}.
\end{equation}
If $w_B$ is the equation of state parameter for the background fluid,
then
tracking behavior with $w < w_B$ occurs when $\Gamma$ is roughly constant
with $\Gamma > 1$ \cite{zlatev}.
Note that $\Gamma$ is exactly constant for power law and exponential
potentials.  In particular, for the potential $V \propto \phi^{-\alpha}$,
we have $\Gamma = 1 + 1/\alpha$.
Taking the derivative
of equation (\ref{motion1}) with respect to $\phi$ gives
\begin{eqnarray}
\Gamma &=& 1 - \frac{2 X^{\prime\prime}}{(1+w)(6 + X^\prime)^2}
- \frac{1-w}{2(1+w)} \frac{X^\prime}{6+ X^\prime}\nonumber\\
\label{motion2}
&+& \frac{3(w_B - w)}{1+w} \frac{1- \Omega_\phi}{6 + X^\prime}.
\end{eqnarray} 
Note that
this equation differs from the corresponding equation
in Ref. \cite{zlatev} because we do not take the limit $\Omega_\phi 
\rightarrow 0$; all of the interesting evolution occurs when
$\rho_\phi$ becomes significant compared to $\rho_M$.  It does
however, agree with the expression previously derived
in Ref. \cite{rubano}, and in the
limit where $\Omega_\phi \rightarrow 0$, equation (\ref{motion2})
reduces to the corresponding equation in Ref. \cite{zlatev}.

When the tracker solution is reached, and the background
fluid dominates, $w$ is a constant.  We will be interested in
the late-time evolution of the quintessence field, so we assume
that the background fluid is cold dark matter, with $w_B = 0$.
Then for the tracker solution, taking $X^{\prime\prime}=
X^\prime = 0$  and $\Omega_\phi=0$ in equation (\ref{motion2}) gives
\begin{equation}
\label{wtrack}
w = \frac{-2(\Gamma-1)}{2(\Gamma-1)+1}
\end{equation}
In this regime, $\Omega_\phi$ increases with time;
when $\Omega_\phi$ is no longer negligible with respect to $1$,
the tracker solution no longer applies.  At this point,
$w$ decreases, asymptotically approaching $w = -1$. (See Ref. \cite{Watson}
for a discussion of the evolution in this regime).

Since $w^\prime \le 0$ for these tracker solutions, we have
$X^\prime \le 0$.  Initially, $X^\prime =0$, but as
$\Omega_\phi$ becomes nonnegligible, $X^\prime$ decreases.
It reaches some minimum value, $X^\prime_{min}$, but then
increases back to a value near zero as $\Omega_\phi \rightarrow 1$.
This behavior is illustrated in Fig. 3, in which
we show the evolution for $X^\prime$ as a function
of $\Omega_\phi$ for several power-law potentials.
\begin{figure}[htb]
\centerline{\epsfxsize=3truein\epsffile{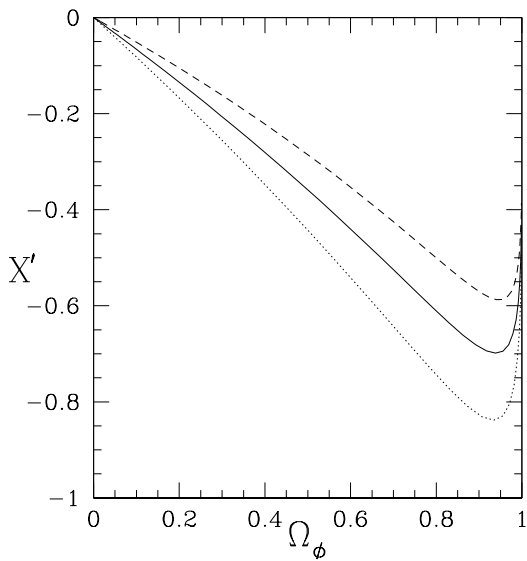}}
\caption{The evolution of $X^\prime$ as a function of $\Omega_\phi$
for the potential $V(\phi) \propto \phi^{-\alpha}$ with
$\alpha = 2$ (dashed curve), $\alpha = 1$ (solid curve) and
$\alpha = 0.2$ (dotted curve).}
\end{figure}
The value of $X^\prime_{min}$ is the key to our calculation,
since it gives a lower bound on $w^\prime$ through equation
(\ref{xprime}).

To find this minimum value for $X^\prime$, we simply take
$X^{\prime\prime}=0$ in equation (\ref{motion2}), where
we assume also that $w_B=0$ and that $\Gamma$ is nearly constant.  This gives
\begin{equation}
\label{xmin}
X^\prime_{min} = -6 \frac{(1-\Omega_\phi)w + 2(\Gamma-1)(1+w)}
{(1-w) + 2(\Gamma-1)(1+w)}.
\end{equation}
On the right-hand side of this equation, $w$, $\Omega_\phi$, and
$\Gamma$ are to be evaluated at the point at which
$X^\prime = X^\prime_{min}$.  In general $\Omega_\phi$ and
$w$ evolve in a complex fashion (see, e.g., Ref. \cite{Watson}),
so equation (\ref{xmin}) does not have a simple solution.  However,
since we seek a lower bound on $X^\prime$, it is sufficient
to use equation (\ref{xmin}) to find a lower bound on $X^\prime_{min}$.
We first note that $\Omega_\phi < 1$; applying this
limit (with $w<0$) to equation (\ref{xmin}) yields
\begin{equation}
\label{xmin2}
X^\prime_{min} > -6 \frac{2(\Gamma-1)(1+w)}
{(1-w) + 2(\Gamma-1)(1+w)}.
\end{equation}
The right-hand side of equation (\ref{xmin2}) increases with
decreasing $w$, so a lower bound on the right-hand side is achieved
by assigning the maximum allowed value to $w$.  In tracker
models, $w$ decreases from the initial value given by
equation (\ref{wtrack}), so substituting the value for
$w$ given by equation (\ref{wtrack}) into equation (\ref{xmin2}) gives
a lower bound on $X^\prime_{min}$:
\begin{equation}
X^\prime_{min} > - \frac{12(\Gamma-1)}{6(\Gamma-1)+1}.
\end{equation}
Taking $X^\prime > X^\prime_{min}$, and using equation (\ref{xprime})
to convert from $X^\prime$ to $w^\prime$, we obtain the limit
\begin{equation}
w^\prime > - \frac{6(\Gamma-1)}{6(\Gamma-1)+1}(1-w)(1+w).
\end{equation}
This is obviously a model-dependent result, since it depends on the
value of $\Gamma$ for which tracking behavior is achieved.  However,
we can use it to derive a more general limit by noting that,
since $\Gamma >1$, we have
$6(\Gamma-1)/[6(\Gamma-1)+1] < 1$, giving
\begin{equation}
\label{wprime2}
w^\prime > -(1-w)(1+w).
\end{equation}
Equation (\ref{wprime2}) gives a lower bound on $w^\prime$ for
tracker models.  This bound is shown in Fig. 2.  This is clearly
a very conservative limit; for example, it is obvious from Fig. 3
that the value of $X^\prime_{min}$ is
often obtained for a value of $\Omega_\phi$ larger than
is consistent with observations.  However, equation (\ref{wprime2})
does give a tighter bound than that derived in Ref. \cite{CL},
further separating quintessence models from the barotropic fluid
models discussed in the previous section.

\section{Discussion}

Our results illustrate the usefulness of the $w - w^\prime$ plane
as a means for observationally distinguishing dark energy models.
Non-phantom barotropic fluid models with positive squared sound speed and
``tracker" quintessence models occupy disjoint regions in this
plane.  Even our less stringent general limit on quintessence models with
monotonic potentials is inconsistent with a wide class of
barotropic fluid models, as can be seen by comparing Figs. 1 and 2.

Supernova Ia measurements \cite{Knop,Riess} have narrowed the range
of possible values for
$w$; in particular, they are consistent with $w \approx -1$.  The value of $w^\prime$
is less well constrained (see, e.g., Ref. \cite{Riess}).  The results
of Ref. \cite{CL} indicate that the resolution on a measurement
of $w^\prime$ must be of order $1+w$ in order to distinguish ``freezing"
and ``thawing" quintessence models.  We reach a similar conclusion
with regard to barotropic fluids and quintessence models: observationally
distinguishing between
these two classes of models
also requires that $\sigma(w^\prime) \lesssim 1+w$.

Finally, we note that there is a relation between the $w - w^\prime$ parametrization
of Ref. \cite{CL} and the ``statefinder" pair introduced
in Ref. \cite{state}.  The statefinders $r$ and $s$ are defined by
\begin{eqnarray}
r &=& \frac{\dddot{a}}{aH^3},\\
s &=& \frac{r-1}{3(q-1/2)},
\end{eqnarray}
where $q = -a \ddot a/\dot a^2$
is the deceleration parameter.  Then $r$ and $s$ are related
to $w$ and $w^\prime$ via \cite{state}
\begin{eqnarray}
\label{r}
r &=& 1 + \frac{9}{2} \Omega_{DE} w(1+w) - \frac{3}{2} \Omega_{DE} w^\prime,\\
s &=& 1 + w - \frac{w^\prime}{3w},
\end{eqnarray}
where $\Omega_{DE}$ is the fraction of the total energy
density contributed by the dark energy.  Note, however, that the mapping between $w$, $w^\prime$
and $r$, $s$ is non-trivial because of the factor
of $\Omega_{DE}$ in equation (\ref{r}).  For non-constant $w$,
the evolution of $\Omega_{DE}$ depends in a complicated way
on both $a$ and $w(a)$.  Hence, the statefinder parameters
encode somewhat different information about the behavior
of the dark energy than do $w$ and $w^\prime$.
The statefinder parameters are more directly related to
observable quantities, while the $w-w^\prime$ parametrization
describes more directly the physical properties of the dark energy.

\acknowledgments

I thank A. A. Sen, R. Caldwell, and E. Linder for helpful discussions.
I am grateful to E. Babichev and C. Rubano for helpful comments
on the manuscript.
R.J.S. was supported in part by the Department of Energy (DE-FG05-85ER40226).

\end{document}